# Conformational variability in proteins bound to single-stranded DNA: a new benchmark for new docking perspectives

*Protein-ssDNA benchmark and its analysis*

MIAS-LUCQUIN Dominique[1*], CHAUVOT DE BEAUCHENE Isaure[1]

[1] Universite de Lorraine, CNRS, Inria, LORIA, Nancy, France

* *dominique.mias-lucquin@loria.fr*

**ABSTRACT**
We explored the Protein Data-Bank (PDB) to collect protein-ssDNA structures and create a multi-conformational docking benchmark including both bound and unbound protein structures. Due to ssDNA high flexibility when not bound, no ssDNA unbound structure is included in the benchmark. For the 91 sequence-identity groups identified as bound-unbound structures of the same protein, we studied the conformational changes in the protein induced by the ssDNA binding. Moreover, based on several bound or unbound protein structures in some groups, we also assessed the intrinsic conformational variability in either bound or unbound conditions, and compared it to the supposedly binding-induced modifications. To illustrate a use case of this benchmark, we performed docking experiments using ATTRACT docking software. This benchmark is, to our knowledge, the first one made to peruse available structures of ssDNA-protein interactions to such an extent, aiming to improve computational docking tools dedicated to this kind of molecular interactions.

## Introduction

While originally described by Watson and Crick[1] as a double helix, composed of two strands bonded together by hydrogen-bonds, DNA is often found in a transient single-stranded state (ssDNA) during its processing, such as genome replication[2], or horizontal gene transfer[3], and bound to proteins. These complexes (ribosomes[4], ICE-relaxase[5], replication fork complex[6], …) are potential therapeutic targets in diseases[7,8].

The structural analysis of these complexes can help to understand how they achieve their function[9]. For example in can reveal the conformational changes undergone by the protein during nucleic acids (NA) binding, by comparing protein structures with and without bound NA[10].

While very informative, high resolution experimental structures of ssNA-protein complexes are expensive and may be difficult, or even impossible, to obtain, due to the inherent poor ordering of NA, especially ssNA[11,12]. Several software systems have tried to implement accurate ssRNA-protein docking, including:

- ATTRACT[13] uses a fragment-based approach, with the need of some knowledge about some protein-RNA contacts;
- RNP-denovo[14], based on Rosetta[15], performs folding and docking of the RNA on the protein simultaneously, but requiring the exact coordinates of few nucleotides;
- RNA-lim[16], models a rough coarse-grained RNA structure (one non-oriented bead per nucleotide) restrained by a set of known binding-sites on the protein surface;

While all these methods advertise a prediction precision from 2 to 10 Å of RMSD between predicted and experimental ssRNA location, none of them was tested yet on ssDNA-protein docking. To our knowledge, no benchmark is available for ssDNA-protein docking. Moreover, while it is possible to query ssDNA-protein complexes with the Nucleic Acid Database[17] (NDB), it seems to find none after 2013, thus limiting the scope of a NDB-derived benchmark. In turn, docking algorithms need experimental ground truth to validate and compare methods. Thus, docking benchmarks based on experimentally resolved structures of complexes are needed. Such docking benchmarks exist for protein-protein[18], membrane protein-protein[19], protein-RNA[20,21], and dsDNA-protein[22] complexes. And while some works studied ssDNA-protein interactions from few structures in the PDB[23], none seems to be as exhaustive as possible, with a primary goal to improve ssDNA-Protein docking.

Here, we present a ssDNA-protein docking benchmark based on structures extracted from the PDB, that contains 91 sequence-identity groups of bound-unbound protein chains, created to evaluate ssDNA-protein docking. Due to the high flexibility of unstructured ssNA, it is not relevant to use their unbound forms in the context of macromolecular docking. This is also the reason why the docking programs presented earlier do not require a known unbound ssRNA structure. In consequence, the main aim of this dataset is to provide bound and unbound structures of the proteins but only bound structures of ssDNA, from ssDNA-protein complexes. When possible, we provide several structures for both bound and unbound states, allowing to differentiate binding-specific from binding-independent conformational changes.

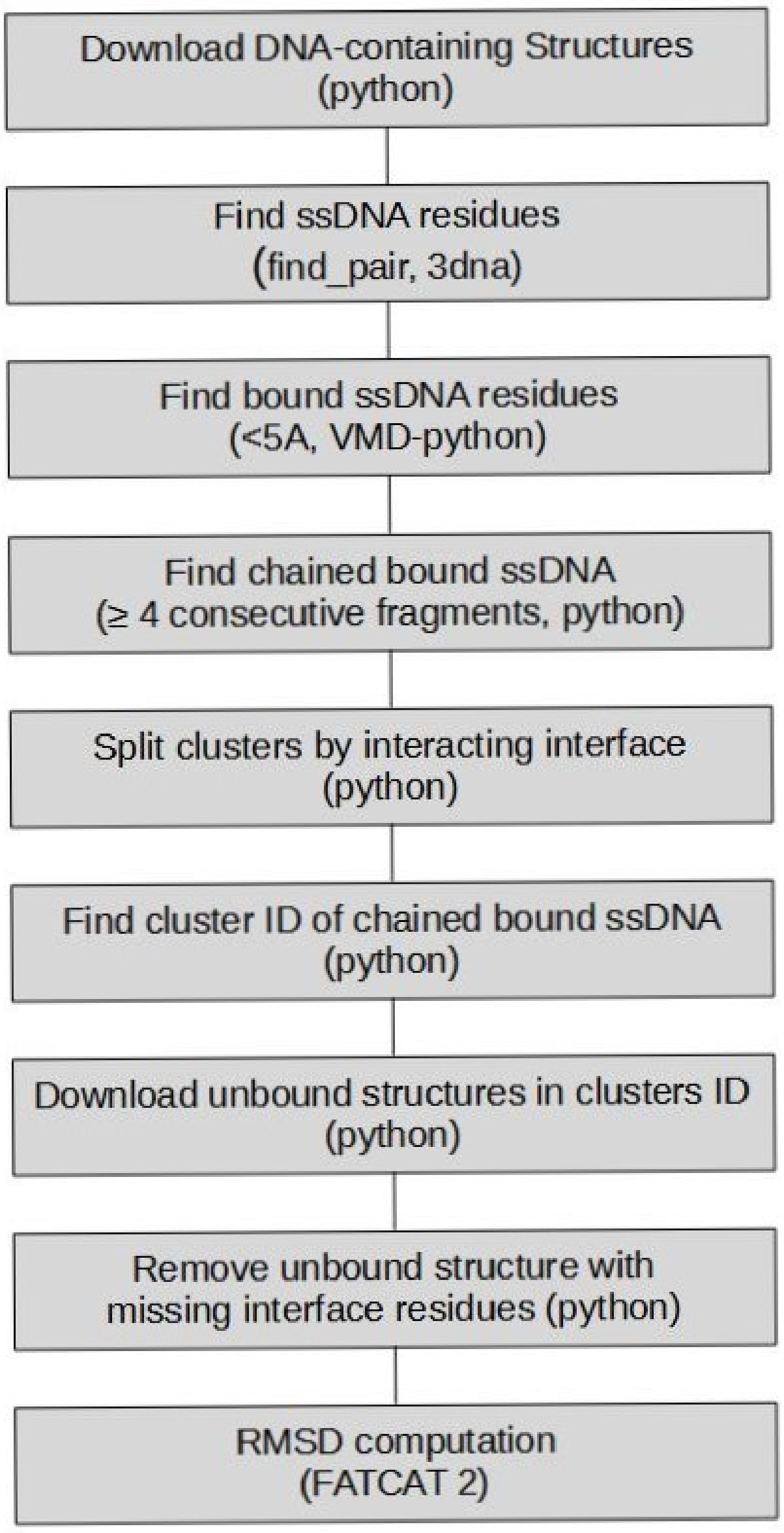

*Fig.1: Protocol to create the benchark*

Docking experiments were performed to show a use case for this benchmark. It underlines the relevance of using several bound structures as ground truth and to tolerate a minimum conformational deviation from ground truth when evaluating docking results.

## Material and methods

All analysis were performed using Python 3.7. Databases were queried on August 18th 2021. Processing steps are summarized in Figure 1.

### *RCSB Protein Data Bank (PDB) query*

We identified the structures containing simultaneously DNA (non hybrid) and proteins by querying the RCSB PDB[24] using their search (https://search.rcsb.org) and data (https://data.rcsb.org) APIs. Another query was performed to extract the PDB ID of all structures containing only proteins without DNA.

These two PDB ID lists were compared to the weekly 100% sequence identity clustering (“clusterNumber100”) of protein chains in the PDB (ftp://resources.rcsb.org/sequence/clusters/) to extract identity groups containing chains being part of both DNA-protein and protein-only structures.

### *Structure alignment, processing and identification of Interacting ssDNA*

For each structure containing DNA and protein, the asymmetric units and all biological assemblies (if any) were downloaded from the PDB[24]. We only kept “ATOM” records describing atoms belonging to proteins or nucleic acids. Nucleic residues involved in double strands are located with 3DNA (script find_pair[25]) in the asymmetric unit and in each biological assembly. A DNA residue is considered single-stranded only if it is not found as double-stranded in any of these structures. Biological assemblies allow the identification of cases where a double strand is formed by the repetition of the asymmetric unit (such as PDB ID:3HZI), while the asymmetric unit eases the processing if the assembly is constituted by the repetition of chains having the same identifier (also like in PDB ID:3HZI). Then VMD-python (Humphrey et al., 1996, https://github.com/Eigenstate/vmd-python) was used to compute distances between ssDNA nucleotides and protein residues; ssDNA nucleotides are bound if they are found at less than 5 Å from any protein residue. This bound ssDNA list is processed to only keep the protein chains interacting with a DNA chain of at least 4 consecutive bound and single-stranded nucleotides. For multi-framed asymmetric units (often encountered with NMR models), only the first frame is used. In this case, we assumeda limited variation of conformation between frames, with no impact on DNA 2D state.

Bound protein chains were then sub-grouped by interaction interface: interaction interfaces with ssDNA are computed with VMD-python; two bound chains in a sequence identity cluster are grouped if they share at least one interacting residues. This can lead to one sequence cluster being split into several if distinct interfaces are found. B

From the RCSB PDB sequence clustering, we retrieved protein unbound chains belonging to structures without DNA that have 100% of sequence identity with previously identified bound chains. Unbound chains were compared to their bound counterpart to identify any structure with missing residues at the interacting interface. If one is found, it is removed from the unbound chains list.

In each sequence identity cluster, bound and unbound chains were rigidly superimposed and global RMSD computed with FATCAT 2.0 standalone software[25]. FATCAT was used because of its ability to superimpose structures with some minor differences between sequences (like missing loop or mutation).

### *Superposition, RMSD calculation and clustering*

The final benchmark is reported in supplementary Table S1 (and RMSD tables in supplementary Table S2), in which chains with a RMSD lower than 0.2 Å are grouped under a single representative chain. This was done to limit the bias from structures containing several times the same chains. These non-redundant dataset is further analyzed. The redundant benchmark and RMSD tables can be found as supplementary Tables 3 and 4 respectively. The sequence identity between clusters is reported in supplementary Table 5.

Interacting ssDNA fragments are reported in supplementary Table 6. We called “fragment” each interacting ssDNA region, while “sequence” refers to unique ssDNA sequence among all clusters. Thus, a DNA chain can have several ssDNA fragments if separated by non-interacting or double-stranded DNA, and several fragments can have the same sequence.

All docking files are available on https://github.com/DomML/ssDNAbenchmark. Because structures can easily be retrieved from the PDB, they were not included in the repository.

### *Docking and docking evaluation*

Docking experiments were performed using ATTRACT[26] without explicit restraints. ATTRACT docking being rigid, docking was performed using a library of multiple tri-nucleotides (reported in Tables 2 and 3) built using protNAff (https://github.com/isaureCdB/ProtNAff). Tri-nucleotides libraries were used as an exhaustive (at 1Å of heavy-atoms RMSD) ensemble of ligand conformations. For each docking, a random selection of 1000000 initial poses was performed, a pose corresponding to one fragment conformation from the library at one random starting position on sphere of 35Å radius around the protein with one random orientation. The fragments position was optimized by gradient descent minimisation of the protein-DNA energy in ATTRACT coarse-grained force field [27].

The best docking solutions according to ATTRACT scoring were analyzed against the bound reference, by measuring the interface RMSD (irmsd), ligand RMSD (lrmsd) and the fraction of native contacts (fnat).

## Results & Discussion

### *Composition of the benchmark*

Here, we present a dataset composed of 284 bound and 669 unbound protein chains, distributed in 91 sequence-identity groups and 98 groups with distinct interfaces: 8 identity clusters contain two distinct interfaces, none have more. It covers a wide range of protein structural families, and should be very useful both for a better understanding of the mechanisms involved in ssDNA-protein binding and as a benchmark to evaluate ssDNA-protein docking software systems.

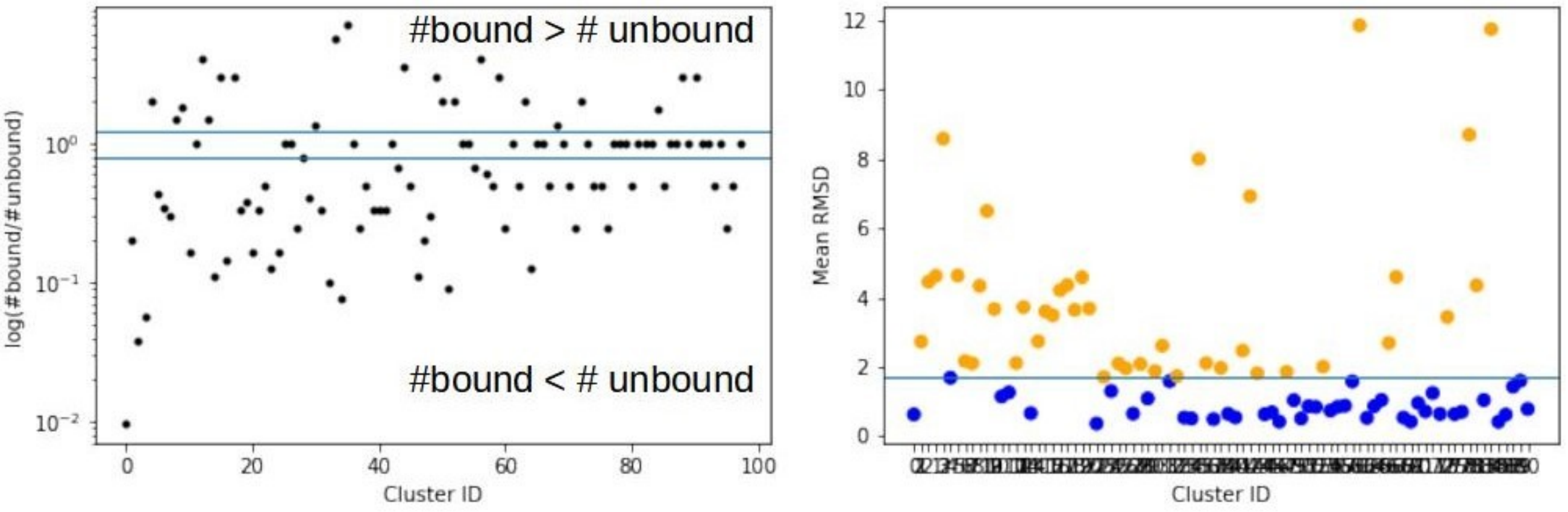

*Fig.2: (A, left) Ratio between number of bound and unbound chains in each cluster, (B, right) Mean-RMSD for the clusters made of at least 3 chains*

The most populated group (group 0, endoribonuclease hydrolase) is composed of 104 chains (1 bound, 103 unbound), and 46 groups have at least two chains for both bound and unbound. These multiple conformations allow the study of conformations variability in the bound or unbound state. Several bound structures also allows the definition of not a unique ground-truth structure, but a range of deviations around one bound structure in which a docking solution is considered as good. Moreover, 8 groups have more than one bound chain but only one unbound chain, 31 groups are in the opposite situation, and 13 contain only one chain for both bound and unbound.

Similarly to group 0, most of the groups are unbalanced: ththey have a difference of at least 20% between the number of bound and unbound chains (Figure 2A).

The presence of several bound or unbound chains for a given group is often purposely avoided in other benchmarks[20–22] by excluding structures with similar sequences. However, offering alternative conformations and alternative docking solutions for a given structure provides essential piece of information to discriminate between conformational changes occurring independently of the presence or absence of bound DNA and changes induced by ssDNA binding. Moreover, as shown by He et al.[28], taking into account pre-binding conformation variability can increase docking performances .

### *Proteins conformational variability*

In each group, the RMSD was computed between each pair of chains to assess the aforementioned conformation variations.

Among the 54 groups with at least 2 bound non-redundant chains (RMSD > 0.2Å), 42 have a mean bound-RMSD of 2 Å or below, marking a general stability of the bound forms. Among the 12 other groups, only 2 have a mean-bound RMSD over 10 Å (#61 and 84).

For the 76 groups with at least 2 unbound non-redundant chains (RMSD > 0.2Å), 48 have a mean unbound-RMSD under 2 Å, 27 between 2 and 10 Å, and 3 over 10 Å: #3, 61 and 84.

We notice a general stiffening of conformations in bound proteins, with a trend towards lower bound-RMSDs than unbound-RMSDs, probably due to ssDNA-binding. This is consistent with what was reported earlier for dsDNA[33].

Overall, in the 85 groups that contain at least three chains, the median mean RMSD (bound + unbound) is 1.69 Å (Figure 2B). This signals a low general variation in conformations between the chains of a group.

Two groups (#61 and 84, Figure S1 and S2) have a mean-RMSD higher than 10 Å. These high RMSD values are due to different relative positions of sub-domains of the protein. Moreover, RMSD clustering shows the presence of several clusters, corresponding to as many major conformational states, but none of them contains all bound or unbound structures. This is a clear indication towards changes that are not driven by DNA binding. While many related works focus on conformational changes upon RNA[29] or DNA[30,31] binding, ligand binding without conformational change has been modeled[32], but poorly studied.

To go further on this track, we dug into the 54 groups containing 2 or more non-redundant bound chains, to test whether there exists a pair of bound-unbound chains closer than a pair of bound-bound chains. If all bound chains are closer to each other than to any unbound chains, this can be a mark of a specific binding-induced rearrangement, including local or low amplitude changes. This is the case for group #3, with a maximum RMSD between bound chains of 3,1 Å and a minimum RMSD between bound and unbound chains of 10.7 Å. Overall, 35 groups are in such a situation, in agreement with what has been observed for RNA-protein binding[33]. This marks an ssDNA-induced fit, and the use of unbound conformations in a docking procedure without taking into account such an induced change can only lead to results of limited accuracy. In the remaining 19 groups, some bound chains were closer to unbound chains than to other bound chains, indicating that conformational changes induced by DNA binding, if any, are of lower amplitude than the intrinsic variability of the protein, like in group #2.2. This points to a predominant conformational selection effect in ssDNA-binding. In these cases, using all unbound conformations, if possible with some additional conformational sampling, for docking should increase the chances to reach a close-to-bound conformation, and improve docking results.

Interestingly, our benchmark could provide the necessary data to investigate which features, other than bound-unbound comparison, could indicate in which category (binding-specific or binding-independent conformational changes) a protein lies. Moreover, the fact that most conformational changes are induced by the ssDNA binding is in agreement with the better results obtained in general (not specifically with ssDNA) by flexible docking methods[34].

### *Length and sequence of bound ssDNA*

In our benchmark, 325 ssDNA fragments are found bound to proteins, with 148 unique sequences (later called just “sequences”). Among these fragments, 34% (110) are homopolymers: 98 contain only “T”, among which 36 and 24 are respectively 4 and 5 nucleotide long; 2 contains only A, 9 only C and 1 only G (Table 1). The 4- and 5-mers poly-T are also the most represented fragments (supplementary Table 7).

While nucleic-acid base composition is highly dependent on the experimental setup, we can still notice an over representation of T-containing sequences and fragments, with respectively 96 and 255 of them where the thymine is the most common base. This is in agreement with the observation that AT-rich sequences are more prone to form single strands than CG-rich sequences, since the A-T base-pairing is weaker than the CG base-pairing[35].

Besides, 30 fragments (corresponding to 24 different sequences) contain non-canonical nucleotides. While such fragments may not be suitable for general purpose docking methods, we keep them to leave the choice to the end user, as some non-canonical residues have a canonical counterpart.

Finally, 116 interacting fragments (56 sequences) have the minimum required size that we chose for our dataset (4 nt), and 258 (111 sequences) are 6 nt or shorter. Overall, the mean size is 5.6 nt and 5.5 nt for sequences and fragments respectively. Only 22 sequences (40 fragments) have more than 7 nts, which may limit the interest of this benchmark for long sequence docking.

### *Docking*

To demonstrate a use case, we performed two docking experiments of three ssDNA fragments on two unbound protein structures from our benchmark, using ATTRACT[26] docking software. Results are summarized in Table 2 and 3. Cluster #4 and 8.1 were chosen, with four bound references used. ATTRACT was chosen for its common use in litterature and its ability to process DNA and to perform ensemble docking.

*Table 1: Occurences of DNA sequences*

| Sequence | Count |
|---|---|
| AAAA | 1 |
| CCCC | 2 |
| CCCCCC | 1 |
| CCCCCCCC | 2 |
| TTTT | 16 |
| TTTTT | 10 |
| TTTTTT | 10 |
| TTTTTTT | 2 |
| TTTTTTTT | 1 |
| TTTTTTTTT | 5 |

In the two experiments, results quality show great variations depending on the unbound structure and ssDNA fragment .

For cluster #4, GAG ssDNA fragment is poorly docked on the two unbound structures, with null fnat values, while GCT fragment is correctly docked on both structures, with low interface RMSD and fnat values over 0.7. On the other hand, fragment AGC is well docked on 1smy_c, with fnat over 0,3, but not on 5tmf_c, where the interaction interface is not found. For cluster #8.1, none of the docked fragments found the interacting interface.

The relevance of redundancy in the benchmark is also clearly shown in our docking experiments. Indeed, in both experiments, bound structures were selected because the same ssDNA sequence was found interacting with the protein, and protein chains were clustered because of the common ssDNA-interface. However, we can observe small variations between the comparison of one docking result with its four bound references, like in the cluster #4, GCT docking on 5tmf_c exhibits irmsd between 4,2Å and 4,4Å depending on the reference. This can be used to set a minimal rmsd clustering criterion of the resuting poses afterdocking experiment, as here 0.2Å in cluster #4.

*Table 2, 3: Docking results*

| | Unbound | 1smy_c | | | | 5tmf_c | | | |
|---|---|---|---|---|---|---|---|---|---|
| | Bound | 4oip_h | 4oiq_h | 4g7h_r | 4q4z_h | 4oip_h | 4oiq_h | 4g7h_r | 4q4z_h |
| GAG (678) | irmsd | 8.195 | 8.333 | 8.176 | 8.164 | 7.673 | 7.722 | 7.764 | 7.756 |
| | lrmsd | 20.775 | 21.011 | 20.950 | 20.934 | 19.774 | 20.041 | 20.035 | 19.962 |
| | fnat | 0.00 | 0.00 | 0.00 | 0.00 | 0.00 | 0.00 | 0.00 | 0.00 |
| AGC (497) | irmsd | 4.408 | 4.633 | 4.600 | 4.601 | 10.436 | 10.741 | 10.654 | 10.542 |
| | lrmsd | 12.023 | 12.080 | 12.230 | 12.120 | 28.732 | 28.898 | 29.216 | 29.028 |
| | fnat | 0.36 | 0.33 | 0.31 | 0.31 | 0.00 | 0.00 | 0.00 | 0.00 |
| GCT (490) | irmsd | 4.002 | 4.007 | 3.999 | 4.017 | 4.199 | 4.323 | 4.389 | 4.333 |
| | lrmsd | 11.688 | 11.679 | 11.639 | 11.696 | 11.380 | 11.547 | 11.539 | 11.527 |
| | fnat | 0.25 | 0.19 | 0.17 | 0.17 | 0.27 | 0.33 | 0.30 | 0.25 |
| | Unbound | 3wod_f | | | | 5xj0_f | | | |
| | Bound | 4g7z_h | 4oir_h | 4q4z_h | 4oio_h | 4g7z_h | 4oir_h | 4q4z_h | 4oio_h |
| AAT (497) | irmsd | 9.814 | 9.816 | 9.758 | 9.800 | 9.868 | 9.707 | 9.727 | 9.809 |
| | lrmsd | 30.346 | 30.315 | 30.256 | 30.328 | 27.727 | 27.702 | 27.637 | 27.712 |
| | fnat | 0.00 | 0.00 | 0.00 | 0.00 | 0.00 | 0.00 | 0.00 | 0.00 |
| ATG (590) | irmsd | 11.584 | 11.575 | 11.512 | 11.523 | 12.128 | 12.159 | 12.120 | 12.213 |
| | lrmsd | 33.453 | 33.474 | 33.461 | 33.662 | 31.797 | 31.802 | 31.788 | 32.010 |
| | fnat | 0.00 | 0.00 | 0.00 | 0.00 | 0.00 | 0.00 | 0.00 | 0.00 |
| TGG (612) | irmsd | 12.987 | 12.958 | 12.891 | 12.910 | 14.018 | 14.028 | 14.038 | 13.898 |
| | lrmsd | 34.938 | 34.933 | 34.955 | 35.139 | 34.692 | 34.719 | 34.757 | 34.866 |
| | fnat | 0.00 | 0.00 | 0.00 | 0.00 | 0.00 | 0.00 | 0.00 | 0.00 |

## Conclusion

Unlike ssRNA, the single stranded form of DNA is almost only found as an intermediate state in the DNA processing mechanisms (like DNA replication[36]), which may be a reason for the lack of study of ssDNA-containing complexes, and ssDNA-protein interactions, from a structural point of view. Yet, those intermediate states play a crucial role in DNA metabolism, and their interactions with proteins are potential targets for therapeutic inhibitors. For instance, the transmission of anti-microbial resistance genes among bacteria could be fought by targeting the excision or transportation of ICE (Integrative Conjugative Elements) ssDNA[5]. Such projects require the knowledge of the 3D structure of such assemblies for rational drug design. To develop computational methods to model their spatial conformation, it is necessary studying the existing experimental structures of such ssDNA-protein complexes.

In this work, we systematically extracted from the PDB all such structures with at least 4 bound single-stranded DNA nucleotides, together with the corresponding unbound structures of the protein, to evaluate and predict the requirements and potential effectiveness of ssDNA-protein docking. We identified groups of proteins with and without ssDNA-induced fit. An application for this benchmark would be as a training set to develop a machine learning tool identifying *a priori* in which category a protein falls. A docking experiment was also performed to illustrate a usecase of this benchmark.

From a biological point of view, we found that the ssDNA composition is biased towards short fragments and homopolymers, with the last feature representing a third of all protein-bound ssDNA sequences. The low number of retrieved sequences (148) may reflect the general low interest for ssDNA-bound protein structures, and a dataset like the one presented here is an important step in their studies.

To our knowledge, this work is the first attempt to aggregate as exhaustively as possible the ssDNA-protein structures available in the PDB. With 98 groups of multi-conformational bound/unbound proteins, this benchmark is an essential first step to understand the mechanisms involved in ssDNA-protein interactions and develop mature ssDNA docking protocols.

**Supplementary**

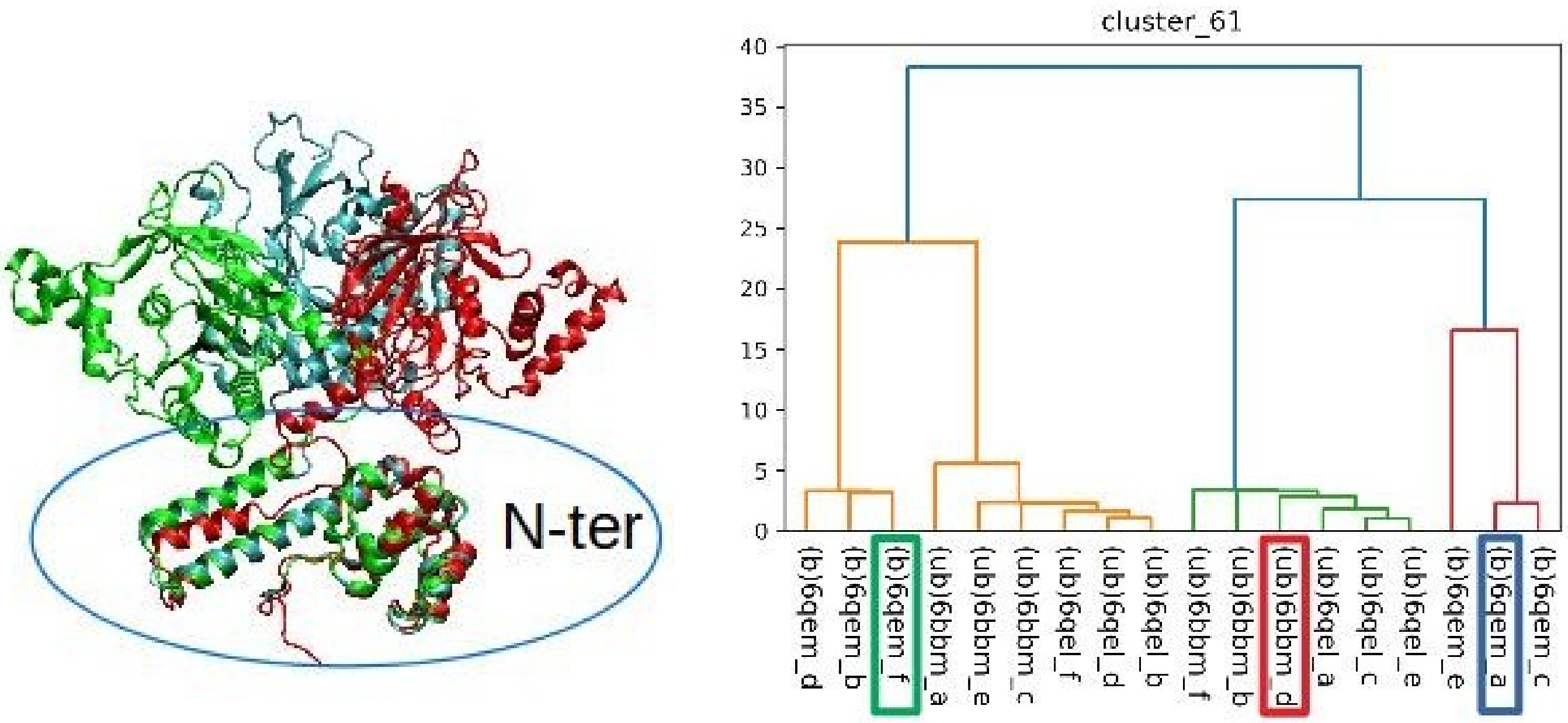

Supplementary figure 1 : A) Superimposition of (blue) 6qem_a, (green) 6qem_f and (red) 6bbm_d by aligning their N-terminal ends ; B) Hierachical clustreing of cluster 61 non-redundant members from RMSD distance matrix ;